\documentclass[fleqn,10pt]{wlscirep}
\usepackage[utf8]{inputenc}
\usepackage[T1]{fontenc}
\usepackage{graphicx,amsmath}
\DeclareUnicodeCharacter{2212}{-}
\title{Tracing the Heliospheric Magnetic Field via Anisotropic Radio-Wave Scattering}

\author[1,*]{Daniel L. Clarkson}
\author[1]{Eduard P. Kontar}
\author[2,1]{Nicolina Chrysaphi}
\author[3]{A. Gordon Emslie}
\author[4]{Natasha L. S. Jeffrey}
\author[5,6]{Vratislav Krupar}
\author[7,8]{Antonio Vecchio}
\affil[1]{School of Physics and Astronomy, University of Glasgow, G12~8QQ, United Kingdom}
\affil[2]{Sorbonne Universit\'{e}, \'{E}cole Polytechnique, Institut Polytechnique de Paris, CNRS, Laboratoire de Physique des Plasmas (LPP), 4~Place Jussieu, 75005~Paris, France}
\affil[3]{Department of Physics \& Astronomy, Western Kentucky University, Bowling Green, KY~42101, USA}
\affil[4]{Department of Mathematics, Physics \& Electrical Engineering, Northumbria University, Newcastle~upon~Tyne, NE1~8ST, United Kingdom}
\affil[5]{Goddard Planetary Heliophysics Institute, University of Maryland, Baltimore County, Baltimore, MD~21250, USA}
\affil[6]{Heliospheric Physics Laboratory, Heliophysics Division, NASA Goddard Space Flight Center, Greenbelt, MD 20771, USA}
\affil[7]{Radboud Radio Lab - Department of Astrophysics, Radboud University, Nijmegen, The Netherlands}
\affil[8]{LESIA, Observatoire de Paris, Universit\'e PSL, CNRS, Sorbonne Universit\'e, Universit\'e de Paris, 5 place Jules Janssen, 92195 Meudon, France}

\affil[*]{daniel.clarkson@glasgow.ac.uk}

\begin{abstract}
Astrophysical radio sources are embedded in turbulent magnetised environments. In the 1~MHz sky, solar radio bursts are the brightest sources, produced by electrons travelling along magnetic field lines from the Sun through the heliosphere. We demonstrate that the magnetic field not only guides the emitting electrons, but also directs radio waves via anisotropic scattering from density irregularities in the magnetised plasma. Using multi-vantage-point type~III solar radio burst observations and anisotropic radio wave propagation simulations, we show that the interplanetary field structure is encoded in the observed radio emission directivity, and that large-scale turbulent channelling of radio waves is present over large distances, even for relatively weak anisotropy in the embedded density fluctuations. Tracing the radio emission at many frequencies (distances), the effects of anisotropic scattering can be disentangled from the electron motion along the interplanetary magnetic field, and the emission source locations are unveiled. Our analysis suggests that magnetic field structures within turbulent media could be reconstructed using radio observations and is found consistent with the Parker field, offering a novel method for remotely diagnosing the large-scale field structure in the heliosphere and other astrophysical plasmas.
\end{abstract}
\begin{document}

\flushbottom
\maketitle

\thispagestyle{empty}

\section*{Introduction}

The majority of astrophysical radio sources are found in turbulent and magnetised environments (e.g., the intergalactic medium, pulsar nebulae, stellar coronae, and planetary magnetospheres). As the emitted radio waves propagate through such background media, they can scatter both elastically and inelastically from ambient density inhomogeneities, resulting in a variety of effects such as shifts in apparent position and angular broadening. Radio wave propagation effects offer not only unique diagnostics (e.g., using plasma dispersion and scattering of fast radio bursts to determine distances \cite{2019A&ARv..27....4P}) but also enormous challenges (e.g., effecting a seeing limitation \cite{1990ARA&A..28..561R}).

Despite decades of observations, the details of particle transport from the Sun through the heliosphere remains a largely open question in space physics \cite{2020A&A...642A...3Z}.  During solar (and stellar) flares, energetic electrons are ubiquitously accelerated \cite{1985ARA&A..23..169D,2017LRSP...14....2B,2024LRSP...21....1K}, 
and some of these electrons escape into interplanetary space 
\cite{2008A&ARv..16....1P,2021NatAs...5..796R,2023ApJ...958...18P,2023ApJ...943L..23K} along the guiding magnetic field lines of the Parker spiral \cite{1958ApJ...128..664P}. These electrons are responsible for the generation of so-called type~III solar radio bursts, in which electromagnetic radiation at radio frequencies is generated via plasma emission processes near the local plasma frequency $f_\mathrm{pe}\,(\mathrm{MHz}) \simeq 8.9 \times 10^{-3} \, n_p^{1/2}$ and/or its harmonic $2f_\mathrm{pe}$ (here $n_p$ (cm$^{-3}$) is the local electron plasma density). As the electrons propagate away from the Sun, at speeds that are a substantial fraction of the speed of light, the local plasma density decreases with distance; thus the observed radio emission forms a \emph{dynamic spectrum}, with emission rapidly shifting to progressively lower frequencies \cite{1959SvA.....3..235G,1959IAUS....9..176W}, typically from hundreds of~MHz down to $\sim$$20$~kHz near $1$~au \cite{1974SSRv...16..189L}. Due to the attenuating effects of the Earth's ionosphere, radiation below $\sim$$10$~MHz is accessible only from space \cite{1974SSRv...16..145F,1984SoPh...90..401B}. Solar radio bursts can be observed simultaneously from multiple spacecraft at different vantage points, providing a unique opportunity to study their propagation through the heliosphere.

Radio waves propagating through a turbulent medium suffer significant scattering due to the spatial variation of the refractive index $n_{\rm ref}=(1 - f_{pe}^2/f^2)^{1/2}$ 
\cite{2019ApJ...884..122K,1952MNRAS.112..475C,1955RSPSA.228..238H,1965BAN....18..111F,1971A&A....10..362S,1999A&A...351.1165A}. Indeed, because solar type~III bursts emit near the plasma frequency, they are particularly susceptible to scattering due to the very low value of the refractive index at such frequencies. Both remote-sensing \cite{2017NatCo...8.1515K, 2020ApJ...905...43C} and spacecraft \cite{2018ApJ...857...82K,2021A&A...656A..34M} observations demonstrate that to explain solar radio burst characteristics, the scattering of radio waves between the source and observer must be taken into consideration. Further, consistency of observed properties with those of the intrinsic sources is achieved only if the scattering process has a substantial degree of wavenumber anisotropy \cite{2019ApJ...884..122K}, with preferential scattering in the direction perpendicular to the magnetic field so that radio waves are preferentially channelled along the magnetic field direction. 
Anisotropic scattering effects in the turbulent plasma mean that triangulating radio sources with multi-spacecraft observations yields only the \emph{apparent} source location, with identification of the location of the \emph{intrinsic} source emission requiring that the effects 
of scattering be disentangled \cite{2021A&A...656A..34M,1989A&A...217..237L,1997SoPh..172..307H,1998JGR...103.1923R,2008A&A...489..419B,2024ApJ...960..101K}. The scattering of radio waves by density fluctuations 
allows sources to be detected by spacecraft at the far side of the Sun, 
albeit with a much weaker observed flux.

Here we compare the results of numerical simulations of anisotropic radio-wave scattering in a turbulent magnetised heliosphere with the $(0.9-0.2)$~MHz observations of solar type~III bursts from four interplanetary radio-observing spacecraft located at different heliocentric longitudes. Our results challenge the traditional assumption that radio waves propagate along straight lines, instead demonstrating that anisotropic scattering significantly alters their trajectory in the presence of turbulent and magnetised plasma. The radio burst intensities received at each spacecraft (from the same source) are consistent with the eastward channelling of radio emission along the spiral geometry of the interplanetary magnetic field lines, and with solar wind speeds of $340-420$~km~s$^{-1}$. Correcting for scattering, the intrinsic locations of the type~III burst sources are found. Comparison of the observations with anisotropic scattering simulations require that the interplanetary field is indeed close to the structure of the Parker spiral, and provides an intriguing new methodology for determining the structure of the interplanetary magnetic field.

\section*{Results}

\subsection*{Observations}\label{sec:observations}

\begin{figure*}[!htb]
    \centering
    \includegraphics[width=0.9\textwidth]{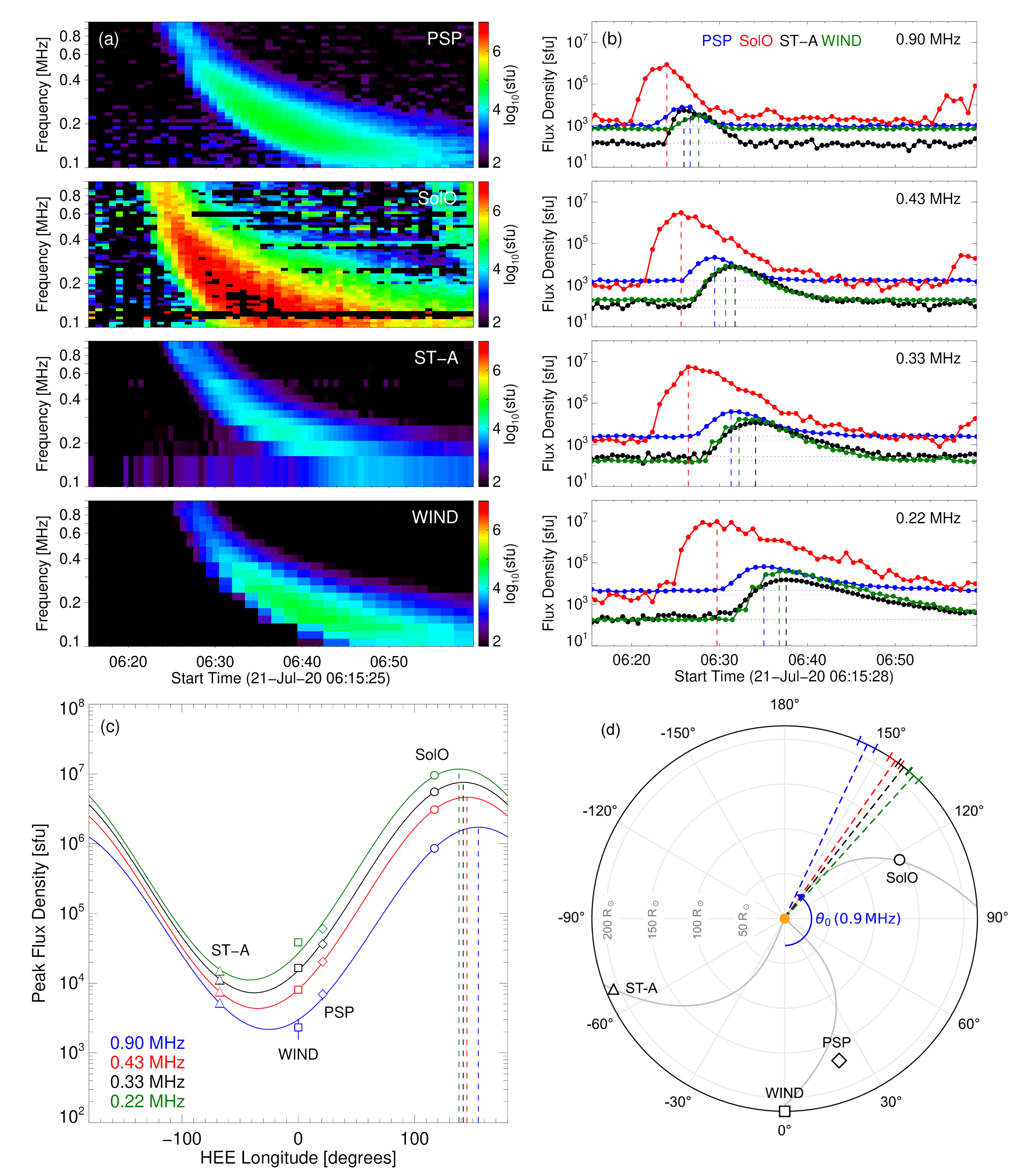}
    \caption{Overview of a type~III burst observed by multiple spacecraft on 2020~July~21 near 06:30~UT. All observed flux densities are scaled to $1$~au. \textit{(a)} Dynamic spectra observed by each spacecraft with the background noise subtracted. \textit{(b)} Time profiles from PSP (blue), SolO (red), ST-A (black), and WIND (green), at select frequencies. The vertical dashed lines denote the observed peak times, and the dotted lines denote the median background level. \textit{(c)} Type~III burst peak intensities (open symbols) observed by each spacecraft at different frequencies. The solid curves show the fits using equation~(\ref{eq:int_fit}); the vertical dashed lines mark the longitude $\theta_0$ of the fitted maximum intensity. \textit{(d)} Longitudes of maximum intensity $\theta_0$ (dashed coloured lines) with colour mapped to the frequencies shown in panel~(c), and the angular uncertainty shown at $1$~au. The blue arc highlights the longitude of maximum intensity $\theta_0$ at $0.9$~MHz. The relative spacecraft positions are shown by the open symbols. The curved grey lines show the possible Parker spiral structure based on the radial solar wind speeds measured at each spacecraft (scaled to $1$~au) of $310$~km~s$^{-1}$ (SolO), $340$~km~s$^{-1}$ (ST-A), and $330$~km~s$^{-1}$ (WIND).}
    \label{fig:ds_lc_peakflux}
\end{figure*}

To observe type~III burst emission from different vantage points around the Sun, we employed radio data from four instruments: Parker Solar Probe/FIELDS (PSP) \cite{2016SSRv..204...49B}, Solar Orbiter/RPW (SolO) \cite{2020A&A...642A..12M}, STEREO A/WAVES (ST-A) \cite{2008SSRv..136..487B,2008SSRv..136..529B}, and WIND/WAVES \cite{1995SSRv...71..231B}. Twenty well-observed type~III burst events were selected, each displaying dynamic spectra at frequencies ranging from 0.9~MHz to below 0.3~MHz (see, for example, Figure~\ref{fig:ds_lc_peakflux}a and Methods).

Figure~\ref{fig:ds_lc_peakflux}b shows the intensity profiles from each probe at approximately 06:30~UT (Earth time) on 2020~July~21, all normalised to a distance of $1$~au using the inverse-square law (see Methods), at four observation frequencies. The heliocentric longitudes of the spacecraft at the time of the observed event are shown in panels~(c) and~(d) of Figure~\ref{fig:ds_lc_peakflux}, from which it is seen that a significant range of longitudes, covering at least half the heliosphere, is represented by the observations. Indeed, at the time of the event, SolO and ST-A were separated by $\sim$$180^\circ$. Figure~\ref{fig:ds_lc_peakflux}b also shows that there is a clear difference in the observed peak times, with the signal arriving first at the spacecraft closest to the Sun (i.e., SolO) that also had the highest normalised intensity flux of all four spacecraft at all observing frequencies. The fact that SolO detected the strongest signal and earliest arrival time suggests that the burst's directivity is aligned most closely with the SolO spacecraft.

The peak flux values (scaled to $1$~au) measured by each spacecraft, $I_{\rm{sc}}$, were fit with the function \cite{2021A&A...656A..34M}
\begin{equation}\label{eq:int_fit}
    I _{\rm{sc}} = I_0 \, \exp{\left(- \, \frac{1-\cos \, (\theta_{\rm{sc}} -\theta_0)}{\Delta \, \mu}\right)} \,\,\, ,
\end{equation}
to determine the direction $\theta_0$ of the maximum flux $I_0$. Here, $\theta_{\rm{sc}}$ is the spacecraft heliocentric longitude, and $\Delta\mu$ is the width of the directivity pattern. In this event, we see a spread in maximum flux of approximately four orders of magnitude around the Sun, peaking near $10^7$~sfu (Figure~\ref{fig:ds_lc_peakflux}c). The direction associated with the peak flux at $0.9$~MHz is near a heliospheric longitude of $154^{\circ}.8$ (where $0^\circ$ is towards the Earth and positive longitudes increase anticlockwise). At lower frequencies, the location of this peak progressively shifts clockwise (eastward), reaching $138^{\circ}.0$ at $0.22$~MHz. We explore changes in $\theta_0$ by defining, at each frequency, the quantity $\Delta\theta$ as the deviation of $\theta_0$ from that at $0.9$~MHz; i.e., $\Delta\theta=\theta_0-\theta_0 \, (0.9~{\rm MHz})$. We then find the average behaviour of $\Delta\theta$ with frequency (as described in Methods), using only events where the best fit has $\chi^2 < 3.84$ ($95$\% confidence for one degree of freedom \cite{1986nras.book.....P}); such a frequency dependence of this longitudinal deviation $\Delta\theta (f)$ is found in all twenty events studied (see Methods for individual events).

\begin{figure}[ht]
   \centering
   \includegraphics[width=0.9\textwidth]{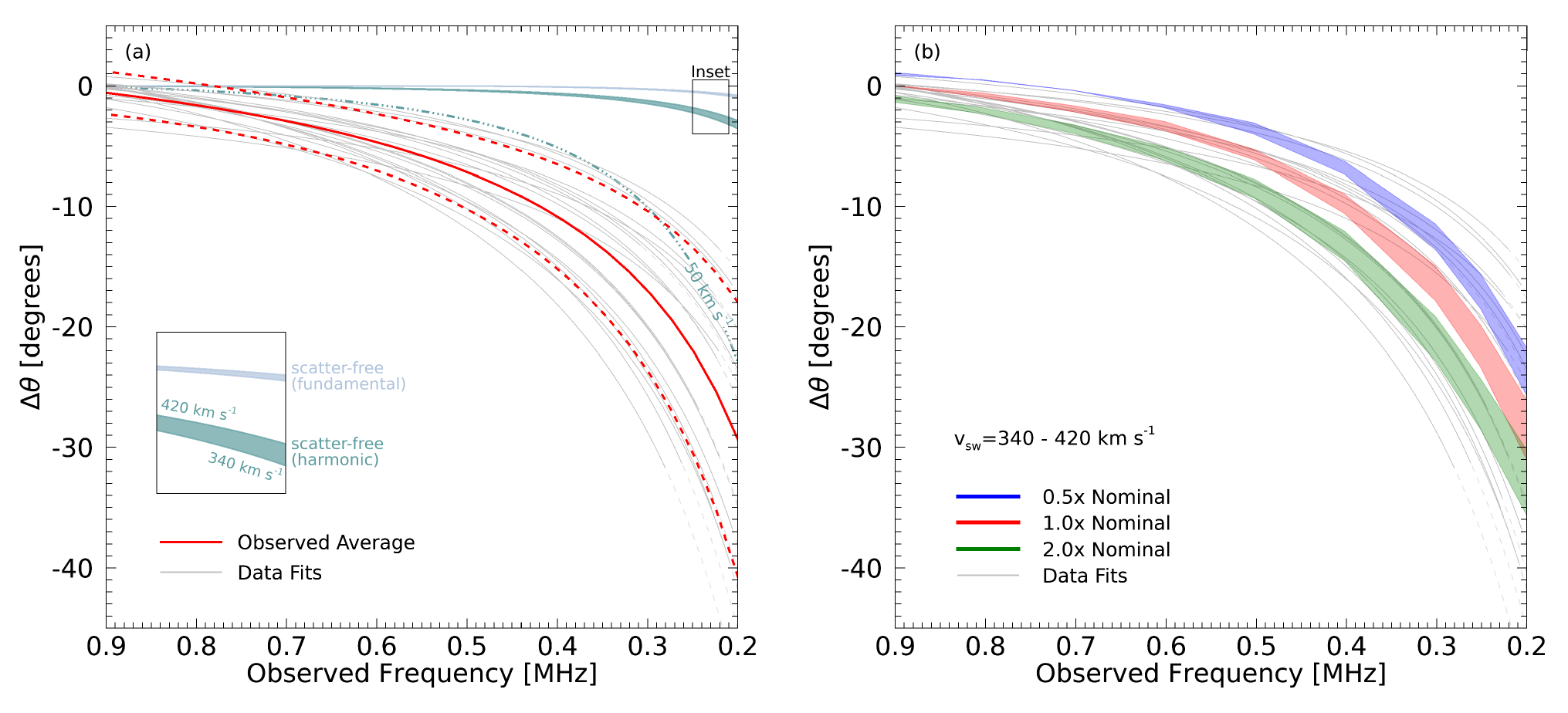}
   \caption{Deviation of the longitude of the fitted peak intensity from that at $0.9$~MHz ($\theta _0$ in Figure \ref{fig:ds_lc_peakflux}). \textit{(a)} Observational data. The solid and dashed red lines show the average and standard deviation of the fitted data from Figure~\ref{fig:angle_shift}. The blue and green bands show the longitude deviation in the scatter-free cases, assuming the emitter propagates along a Parker spiral at distances corresponding to fundamental (blue) and harmonic (green) frequencies. The angle of the Parker spiral at a given distance for a given solar wind speed is calculated as $\theta = -\arctan(\Omega r/v_\mathrm{sw})$. The upper and lower bounds of the coloured bands correspond to a range of $v_\mathrm{sw} (1~{\rm au})$ for the solar wind between 340~km~s$^{−1}$ (lower bound) and 420~km~s$^{−1}$ (upper bound) \cite{2020A&A...635A..44L}, as shown in the inset. The dot-dashed line represents the harmonic scatter-free case with an unrealistic solar wind speed of 50~km s$^{-1}$. \textit{(b)} Simulation data using different turbulence scaling factors between $(0.5-2.0) \, \times$ the nominal profile \cite{2023ApJ...956..112K}. The bounds of each band correspond to different solar wind speeds as in the left panel. The sources are injected along the same Parker spiral field line for a given $v_{\mathrm{sw}}$. For each simulation, the longitude of peak directivity $\theta_0$ is referenced with respect to that at $0.9$~MHz from the nominal simulation data.}
\label{fig:dth}
\end{figure}

Figure~\ref{fig:dth} shows the spread of $\Delta\theta$ from the twenty events based on the fitted data in Figures~\ref{fig:th0_fits} and~\ref{fig:angle_shift}. Fitting the data in frequency space, or with a fundamental or harmonic assumption, produces a consistent result (Figure~\ref{fig:angle_shift})---the mean and standard deviation of the observations at $0.2$~MHz is $(-30 \pm 11)^\circ$. Since the electrons responsible for the type~III burst radio emission propagate along a magnetic field spiral, one might expect a deviation due to different locations of the intrinsic sources at different frequencies (i.e., different ambient densities along the magnetic field). However, the observed longitudinal deviation is too large to be explained by this effect. For the observed longitudinal deviation to be caused by emitting regions being situated at different locations along the Parker spiral, the field lines would have to deviate on average $\sim30^\circ$ between $(0.9-0.2)$~MHz (Figure~\ref{fig:dth}a), and such a large angular deviation of the magnetic field direction over such a relatively small difference in solar distance would require a very slow solar wind speed in the Parker model of $\sim50$~km~s$^{-1}$ (a slower wind speed $v_\mathrm{sw}$ corresponds to a stronger curvature of the magnetic field given by $\theta = -\arctan(\Omega r/v_\mathrm{sw})$, leading to a larger deviation of the apparent source longitudes). Such a low solar wind speed is inconsistent with in-situ measurements of the slow solar wind at the ecliptic that are typically about $400$~km~s$^{-1}$ (see Methods). Hence, if one assumes the radiation propagates in a scatter-free manner and considers only the electron path along the Parker spiral, the angle $\Delta\theta$ would be far too small to be consistent with the observations. Either the typical observed solar wind speeds and magnetic field structure used in the above analysis are fundamentally incorrect, or there must be an additional mechanism that drives the much stronger variation of $1$~au flux with heliocentric longitude. We consider just such a mechanism in the next section.

\subsection*{Modelling: Anisotropic Radio-wave Propagation} 
    
\begin{figure}[t!]
    \centering
    \includegraphics[width=0.9\textwidth]{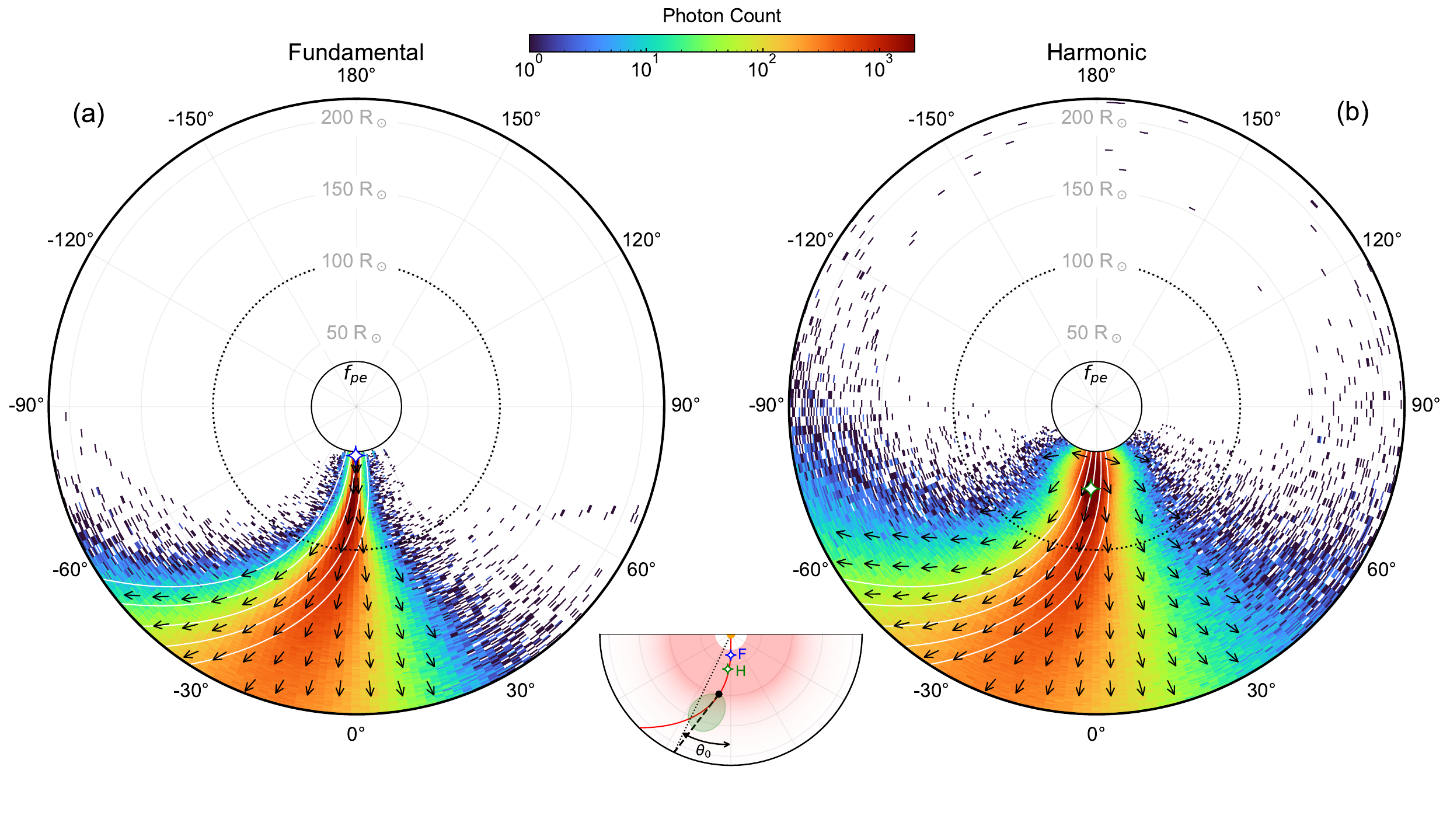}
    \caption{Polar plots of the time-averaged simulated photon propagation in the heliosphere for (a) a fundamental emitter at $r = 34.1 \, R_\odot$ and $\theta = -0^\circ.86$ (open blue star), and (b) a harmonic emitter at $r = 57.4 \, R_\odot$ and $\theta = -3^\circ.2$ (open green star), with anisotropy factors $\alpha=0.25$ and $\alpha=0.4$, respectively. The coloured histograms show the photon positions, and the arrows represent the average wavevector of photons located within grid cells of $10 \, R_\odot \times 10 \, R_\odot$. The white spiral lines show representative Parker field lines for $v_\mathrm{sw}=380$~km~s$^{-1}$, the inner solid black ring denotes the radius of the plasma frequency surface, and the dotted black circle highlights the approximate radius of last scattering at $\sim100 \, R_\odot$, based on Figure~5 of ref. \cite{2023ApJ...956..112K}. The inset visualises the effect of anisotropic radio-wave scattering on the emission pattern for the $0.2$~MHz fundamental and harmonic sources. The red filled region denotes the distances within which strong scattering occurs close to the plasma frequency. The solid red curve shows the Parker spiral along which the type~III burst sources (both fundamental and harmonic) are assumed to propagate. The green lobe shows the resulting emission directivity pattern of the scattered source, centered on the black circle and tangent to the field line near the last scattering radius, with the dashed black line showing the peak direction to $1$~au. The dotted black line gives the radial direction implied by a fit to $1$~au intensity vs. heliocentric longitude $\theta_0$ (equation~(\ref{eq:int_fit})).} 
    \label{fig:summed_photons}
\end{figure}

Radio waves propagating in a turbulent environment experience scattering off density inhomogeneities due to the fluctuating refractive index $n_{\rm ref}$ along the propagation path. Adding a magnetic field (see Methods), which determines the local anisotropy direction, to an anisotropic scattering model \cite{2019ApJ...884..122K}, we traced photons emitted by different injected point sources at frequencies in the range from $0.9$ to $0.2$~MHz, for both fundamental and harmonic emission, out to $1$~au. 
The inset of Figure~\ref{fig:summed_photons} demonstrates this anisotropic scattering effect for both fundamental and harmonic $0.2$~MHz sources and an ambient magnetic field described by a Parker spiral with a solar wind speed of 380~km~s$^{-1}$, the average of the $340-420$~km~s$^{-1}$ range \cite{2020A&A...635A..44L}. Close to the plasma frequency or its second harmonic, the emission is strongly scattered, with the anisotropic density fluctuations acting to channel the emission preferentially along the magnetic field. The scattering rate decreases with heliocentric distance, creating an (approximate) ``surface of last scattering'' beyond which the waves are only weakly scattered. Hence, the direction of peak emission for both the fundamental and harmonic source of the same frequency is approximately tangential to the local magnetic field at the radius of this surface of last scattering (black dashed line in the inset of Figure~\ref{fig:summed_photons}). If we take this surface to be at $r_s = 0.5$~au, then we have $\theta_0 = -45^\circ \, (r_s/1\;\mathrm{au}) \simeq -23^\circ$, in excellent agreement with the average observed value of $\Delta\theta$ presented in Figure~\ref{fig:dth}.

Panels~(a) and~(b) of Figure~\ref{fig:summed_photons} show the aggregate of all photon positions for $0.2$~MHz emission, across all simulation times. Whilst photons are scattered across a wide range of heliocentric longitudes, there is a distinct channeling along the magnetic field direction. This is clearly depicted by the average wavevectors shown in Figure~\ref{fig:summed_photons}, which are tangent to the local field line. Note that there are regions around the heliosphere where, if an observer is sufficiently close to the Sun, they may receive substantially \emph{less} flux than they would at $1$~au, due to the predominant channeling along the direction of the Parker spiral. For example, Figure~\ref{fig:summed_photons} shows that a spacecraft orbiting within $100~R_\odot$ at $\theta = -45^\circ$ would receive a flux level that is orders of magnitude lower than if it were located further away (e.g., close to $1$~au) at the same observation angle.
        
We have also investigated how a change in the scattering rate (incorporated through a scaling factor applied to the turbulence intensity profile) can affect the deviation along the interplanetary field direction. We considered variations over a factor between $(0.5-2) \, \times$ the nominal turbulence profile \cite{2023ApJ...956..112K}. Such a change in the scattering rate leads to a change of radius within which scattering is significant, causing the longitudinal shift of the observed emission to change by about $10$~degrees between the two extremes. We find that the observed gradient of the longitudinal shift of the peak intensity with frequency is always steeper than can be produced in a scatter-free regime by considering electron motion, whilst the wave scattering simulations result in a quantity that is in good agreement with the observations (Figure~\ref{fig:sim_plots}). Thus, the observed longitudinal deviation results predominantly from the anisotropic wave scattering process, with the stronger angular change from the simulations corresponding to those with a high turbulence scaling factor. Additional, yet smaller, variation in angles can be accommodated through solar wind speeds (Figures~\ref{fig:dth} and~\ref{fig:sim_plots}). Varying both the solar wind speed between $340-420$~km~s$^{-1}$, and the scattering rate by a factor of $(0.5-2)$ of the nominal density turbulence level, can reproduce the steeper gradient (Figure~\ref{fig:sim_plots}) of the observed peak drift with frequency, and can also account for the spread in $\Delta\theta$ shown in the observations. 

\begin{figure}[ht]
   \centering
   \includegraphics[width=1\textwidth]{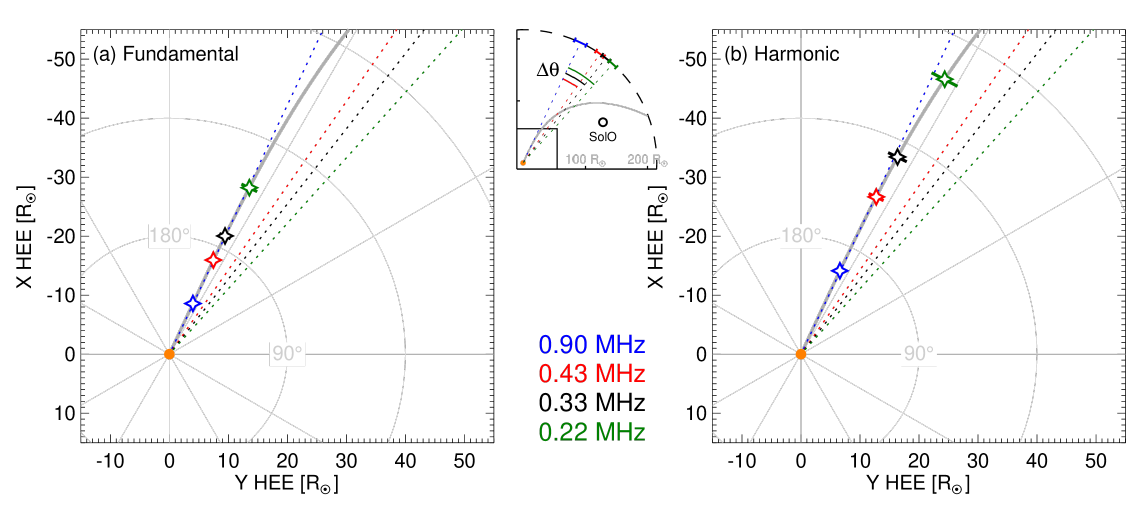}
    \caption{Source emission locations inferred by correcting for radio wave transport for the type~III burst on 2020~July~21, assuming (\textit{a}) fundamental and (\textit{b}) harmonic sources. The source positions of the fundamental ($r=[9.6, 17.7, 22.2, 31.4] \, R_\odot$ and $\theta=-[155.1, 154.9, 154.7, 154.2]$ degrees) and harmonic ($r=[15.8, 29.7, 37.2, 52.6] \, R_\odot$ and $\theta=-\left[155.0, 154.3, 153.7, 151.8\right]$ degrees) are shown by the stars at radial distances inferred from the nominal simulations to the field line that is rooted at the Sun in alignment with the fitted longitude at $0.9$~MHz. The uncertainty in this fitted longitude of $\pm2.7$~degrees is also shown on the source positions. The average $1$~au scaled solar wind speed measured across all spacecraft of $v_{\mathrm{sw}}=310$~km~s$^{-1}$ has been used, as best inferred from Solar Orbiter (Figure~\ref{fig:ds_lc_peakflux}). Varying $v_{\mathrm{sw}}$ weakly affects the source angles: for instance, using $v_{\mathrm{sw}}=380$~km~s$^{-1}$ instead of $v_{\mathrm{sw}}=310$~km~s$^{-1}$ shifts the $0.2$~MHz sources by $-0.2$ degrees (fundamental) and $-0.6$ degrees (harmonic). The inset shows a zoomed out region of the heliosphere with the fitted longitudes $\theta_0$ depicted at $1$~au as described in Figure~\ref{fig:ds_lc_peakflux}. The lower left square in the inset represents the region shown in the main panels.}
\label{fig:emission_sources}
\end{figure}

\section*{Discussion}

Radio emission in astrophysical sources often provides unique diagnostics on the high-energy electrons that emit the radio waves, via a variety of processes such as gyrosynchrotron radiation and plasma emission. The emitting electrons are, of course, guided by the magnetic field so that the spatial distribution of radio wave \emph{sources} is largely governed by the structure of the ambient magnetic field. Here we have used multi-spacecraft observations in conjunction with numerical simulations to demonstrate the largely unappreciated fact that, due to anisotropic scattering of the emitted radio waves off density inhomogeneities in the medium through which they propagate, the emitted radio {\it waves} are preferentially channelled along the direction of the guiding magnetic field, affecting the radiation received by spatially disparate observers. The spread of radio waves is in addition to the spatial spread of type III generating electrons due to an expanding magnetic flux tube.

A combination of multi-spacecraft observations of type~III solar radio bursts observed from multiple vantage points reveals a significant eastward longitudinal displacement 
in the radio wave distribution with decreasing frequency. 
This effect is pronounced both for individual solar radio bursts 
and on average for $20$ type~III bursts. 
Moreover, the same qualitative conclusion holds when 
assuming either harmonic or fundamental emission of the radio bursts. 
From all the events, we find that between $(0.9-0.2)$~MHz, 
the peak direction changes by $(30 \pm 11)^\circ$, 
and the gradient of the apparent source deviation with distance to be $-210^\circ $~au$^{-1}$ for fundamental emission and $-130^\circ $~au$^{-1}$ 
for harmonic emission (see Methods). Both these gradients are far steeper than the typical Parker spiral. Thus, assuming (1) that the magnetic field guides only the emitting electrons whilst (2) the radiation is weakly scattered, cannot explain the directivity pattern in the multi-spacecraft observations; one would have to invoke a much steeper field curvature (corresponding to abnormally slow solar wind) at the distances corresponding to kilometric emission in order to explain the observations through electron motion alone. On the other hand, incorporating anisotropic scattering of the emitted radio waves \emph{does} offer an eloquent explanation of these observations. We observe a consistent eastward shift in the radio emission peak across multiple events, indicating the significant role of anisotropic scattering in shaping the observed emission. Exploiting this longitudinal shift allows one to disentangle the scattering effects to estimate intrinsic source locations as demonstrated for the event of 2020~July~21 in Figure~\ref{fig:emission_sources}. Moreover, comparison of the observations with the simulations allows one to remotely infer the magnetic field structure: the magnetic field from the average of $20$ type~III bursts is found to be close to a Parker spiral with a typical solar wind speed of about $400$~km~s$^{-1}$. A substantially different magnetic field structure or solar wind speeds would be inconsistent with the observations. 

These results demonstrate that magnetic field structures in turbulent astrophysical plasmas can be remotely inferred using radio measurements because of anisotropic scattering of the emission, offering a powerful diagnostic tool for space weather studies. In the case of solar radio burst sources, the presence of the magnetic field causes a pronounced longitudinal variation in the intensity of the observed radio waves, which we can use to trace the field structure. The anisotropy in the emission pattern of radio sources observed at a distance offers the intriguing and potentially wide-ranging diagnostic of the magnetic field structures of different astrophysical environments in which radio sources are embedded.

\section*{Methods}\label{sec:methods}

\subsection*{Multi-spacecraft Type~III Burst Observations}\label{sec:spacecraft}
 
Spacecraft instrumentation measures voltage fluctuations of frequency $f$ induced on a short dipole antenna and hence provides the voltage power spectral density $V^2_f$ (V$^2$~Hz$^{-1}$), a quantity related to the source brightness $B_{\rm{s}}$ (W~m$^{-2}$~Hz$^{-1}$~sr$^{-1}$) by (see Equation~(A1) in ref. \cite{2011RaSc...46.2008Z})

\begin{equation}\label{eq:calib}
V^2_f = \frac{1}{2} \, Z_0 \, \left( \Gamma \, l_{\rm{eff}} \right)^2 \int_{\Delta\Omega_{\rm{s}}} B_{\rm{s}} \sin^2 \psi \, \mathrm{d}\Omega
= \frac{1}{2} \, Z_0 \, \left( \Gamma \, l_{\rm{eff}} \right)^2 \,  I \,\,\, .
\end{equation}
Here $Z_0 = 377 \,\Omega$ is the impedance of free space, $\Gamma \, l_{\rm{eff}}$ is the effective antenna length weighted by a gain factor $\Gamma $, $I=B_{\rm{s}} \, \Delta\Omega_{\rm{s}}$ (W~m$^{-2}$~Hz$^{-1}$) is the flux density of the type~III source located near the Sun and subtending a solid angle $\Delta\Omega_{\rm{s}}$, $\psi$~is the angle between the incoming wave direction and the antenna axis, with $ \sin^2 \psi = 1$ when the incoming radio waves are approximately perpendicular to the antenna direction \cite{2011RaSc...46.2008Z,2021A&A...656A..33V}, 
and the factor of $1/2$ comes from averaging over wave polarisation states. Type~III solar radio burst flux density data are either available from NASA’s Space Physics Data Facility (SPDF), or converted from the measured voltage fluctuations using equation~(\ref{eq:calib}) and the instrument's gain-weighted effective antenna length $\Gamma \, l_{\rm{eff}}$.

\begin{figure}[b!]
  \includegraphics[width=1\textwidth]{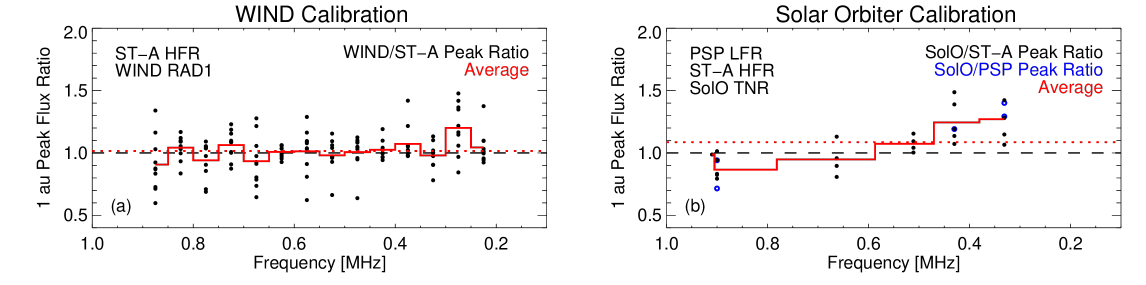}
  \caption{Type~III calibrated peak flux comparison by multiple spacecraft when aligned in heliospheric longitude. \textit{(a)} Ratio of WIND and ST-A calibrated peak fluxes scaled to $1$~au for $10$ type~III burst events that occurred between 2023~August~13--17 (black circles). The red histogram gives the average ratio per frequency, and the dotted red line shows the average ratio across all frequencies of $1.02$. \textit{(b)} As in the left panel, but for SolO ratios against ST-A and PSP during 2021~September~21 (PSP, blue open circles) and 2022~November~30 to 2022~December~03 (ST-A, black circles). The average ratio across all frequencies is $1.09$.}
\label{fig:L2_cal}
\end{figure}

For the STEREO/Waves instrument \cite{2008SSRv..136..487B} onboard ST-A, the L3~data from its High Frequency Receiver (HFR), which operates in the range $0.125-16$~MHz with a bandwidth of $50$~kHz are already available in flux density units (W~m$^{-2}$~Hz$^{-1}$). The RAD1 receiver of the Waves experiment \cite{1995SSRv...71..231B} onboard the WIND spacecraft covers a frequency range from $0.02$ to $1.04$~MHz with a $4$~kHz bandwidth and has a value of $\Gamma \, l_{\rm{eff}}=30$~m. Using the approximate co-spatiality of ST-A and WIND in August~2023, the calibrated $1$-au-scaled peak fluxes from ten type~III bursts observed by both spacecraft between $0.9$ and $0.2$~MHz are compared in Figure~\ref{fig:L2_cal}a. The data show good agreement between the two spacecraft, with an average peak flux ratio of~$1.02$ across all frequencies.

\begin{figure}[t!]
  \centering
  \includegraphics[width=0.8\textwidth]{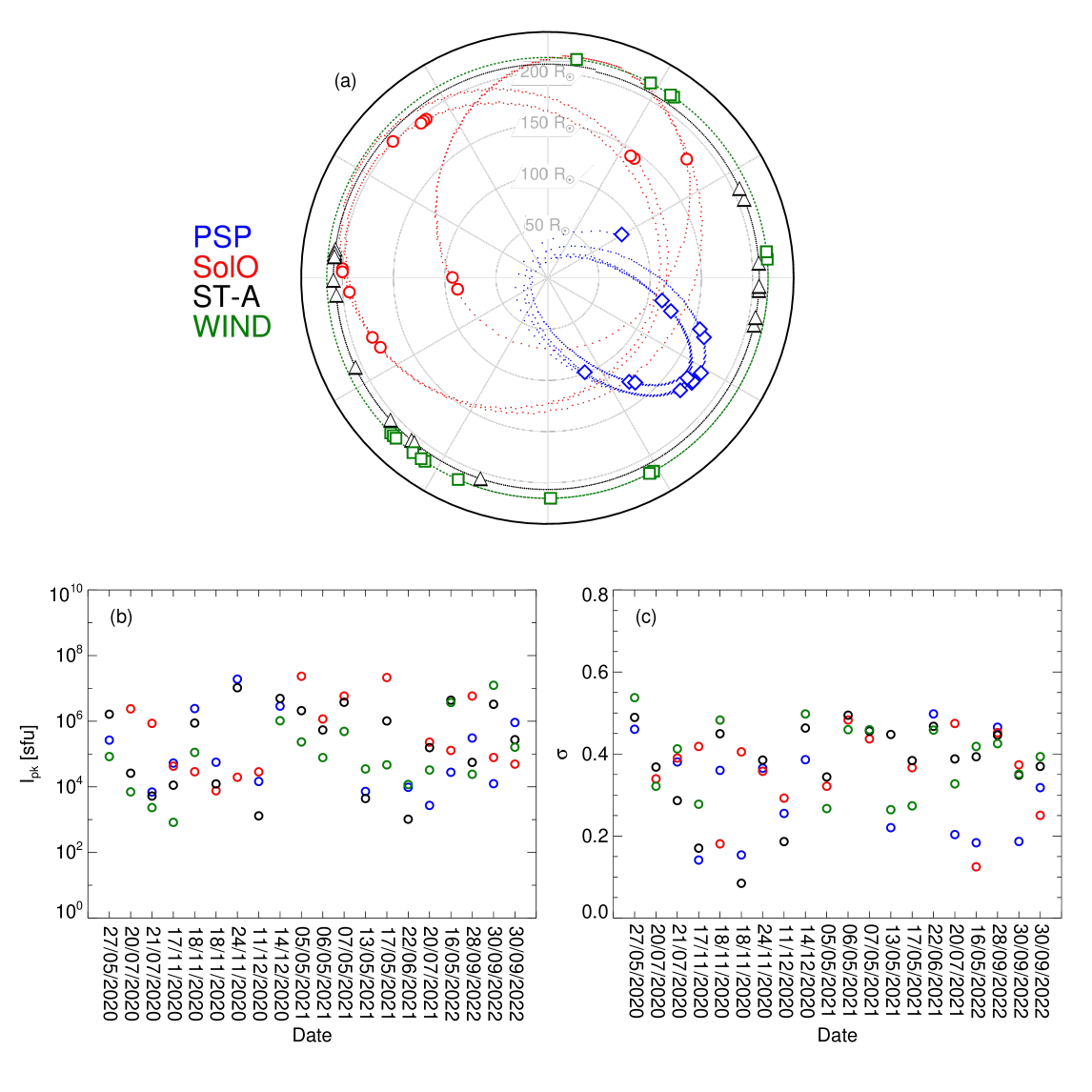}
  \caption{\textit{(a)} Spacecraft positions in the Solar Ecliptic Coordinate System. The orbits (dots) show daily positions between 2020 May 27 and 2022 Sept 30, and the symbols denote the positions of each spacecraft on the dates used. \textit{(b)} Peak flux $I_\mathrm{pk}$ measured by each spacecraft at each frequency, with background subtracted. \textit{(c)} Spread of the measured $I_\mathrm{pk}$ values per spacecraft, where $\sigma$ is the standard deviation of $I_\mathrm{pk}$ at each frequency, normalised by the maximum in each set.}
\label{fig:sc_data}
\end{figure}

For the Radio and Plasma Waves experiment on board Solar Orbiter (RPW/SolO) \cite{2020A&A...642A..12M}, we use the dipole V1--V2 measurements which contain the least noise \cite{2021A&A...656A..33V}. We use the L2~data and calculate the flux density using a value of $\Gamma \, l_{\rm{eff}} = 2.9$~m (see Figure 4 in ref. \cite{2021A&A...656A..33V}). Through comparison of calibrated PSP and ST-A data for eight type~III bursts that occurred when the PSP and ST-A spacecraft were in longitudinal alignment with SolO, we find good agreement between the background subtracted peak fluxes with a ratio ranging between approximately $0.85$ and $1.25$ from $0.9$ to $0.3$~MHz, and an average across all frequencies of $1.09$ (Figure~\ref{fig:L2_cal}b) which is found consistent with that used in \cite{2024ApJ...961...88K}.

Finally for PSP, we use data from the Radio Frequency Spectrometer (RFS) \cite{2017JGRA..122.2836P} onboard the FIELDS instrument \cite{2016SSRv..204...49B}. We use the L3~data from the Low Frequency Receiver (LFR) of RFS that observes at frequencies between $0.1-1.69$~MHz. The frequencies are approximately logarithmically spaced with a constant bandwidth ratio $\Delta{f}/f \simeq 4.5\%$, and the flux density is directly provided in units of W~m$^{-2}$~Hz$^{-1}$.

To ensure that differences in observed intensity are correctly interpreted in terms of directivity effects related to the spacecraft heliocentric longitude $\theta_\mathrm{sc}$, the intensities measured by each probe were scaled from the spacecraft's heliocentric distance $r_{\rm{sc}}$ to $1$~au using the factor $(r_{\rm{sc}}/1\;\rm{au})^2$, and are presented in solar flux units (sfu), where $1$~sfu = $10^{-22}$ W m$^{-2}$ Hz$^{-1}$. We use data from over two years, enabling the spacecraft to cover a wide range of configurations in the heliosphere due to the orbits of PSP and SolO (Figure \ref{fig:sc_data}a). For any given event, they spacecraft have at least $\sim100^\circ$ separation, and all spacecraft locations deviate from the ecliptic plane by only a relatively small angle between $-3^\circ.4$ to $6^\circ$. 

The flux data for each spacecraft is background subtracted by taking the median intensity at each frequency from a quiet period prior to each burst as the background level $\delta{I_{\rm{bg}}}$. For each event, four frequency channels at each spacecraft are selected. Where a specific frequency is not available for a given instrument due to lower spectral resolution, the nearest channel is selected. We avoid frequency channels where there is clear radio interference---in particular, avoiding the electromagnetic contamination visible in SolO/RPW spectra \cite{2021A&A...656A..41M}. As such, the chosen frequency channels are not the same in each event, but cover the range from $0.9$~MHz to below $0.3$~MHz. The peaks of each light-curve $I_{\rm{pk}}$ and their spread measured by each spacecraft per event are shown in panels~(b) and (c) of Figure \ref{fig:sc_data}, and have an uncertainty $\delta{I}$ given by a combination of the signal error and background level as $\delta{I}=\sqrt{(0.12 \, I_{\rm pk})^2 + (\delta I_{\rm bg})^2}$, where the signal error is taken as $12$\% of the peak flux ($\sim$$0.5$~dB amplitude resolution) \cite{2020A&A...642A..12M}, which is in line with the average cross-calibration ratio from SolO of $\sim10$\% across all frequencies (Figure~\ref{fig:L2_cal}b). Across the 20 events, no single spacecraft consistently received systematically the highest or the lowest flux density, suggesting no systematic bias in the heliocentric angles between the emitters and any one spacecraft.

\subsection*{Solar wind speed}\label{sec:solar_wind}

The solar wind speed $v_\mathrm{sw}$ varies at different locations around the Sun, and has been measured using in-situ observations for our twenty events. These measurements provide some insight into the typical spread in $v_\mathrm{sw}$ 
in the heliosphere during each event, allowing us to quantitatively map the Parker spiral structure of the magnetic field. Scaled to $1$~au through the mass continuity equation $4 \pi r^2 \, n \, v_\mathrm{sw} = {\rm constant}$, we obtain
$v_\mathrm{sw} \,
(1~{\rm au}) \simeq v_{{\rm sw, sc}} \, \cdot \,
[n_p(r_\mathrm{sc})/n_p(1\,\mathrm{au})] \cdot
[r_\mathrm{sc}/1\;\mathrm{au}]^2$, and we find a mean and standard deviation for $v_\mathrm{sw} \, (1~{\rm au})$, for all spacecraft and 20 events (see Methods), 
of $(421 \pm 104)$~km~s$^{-1}$, which is a typical slow solar wind speed at $1$~au \cite{2016SSRv..201...55A}. 

\subsection*{Individual events: Source Angular Deviation versus Frequency}\label{sec:ang_dev_freq}

\begin{figure}[b!]
   \centering
   \includegraphics[width=1\textwidth]{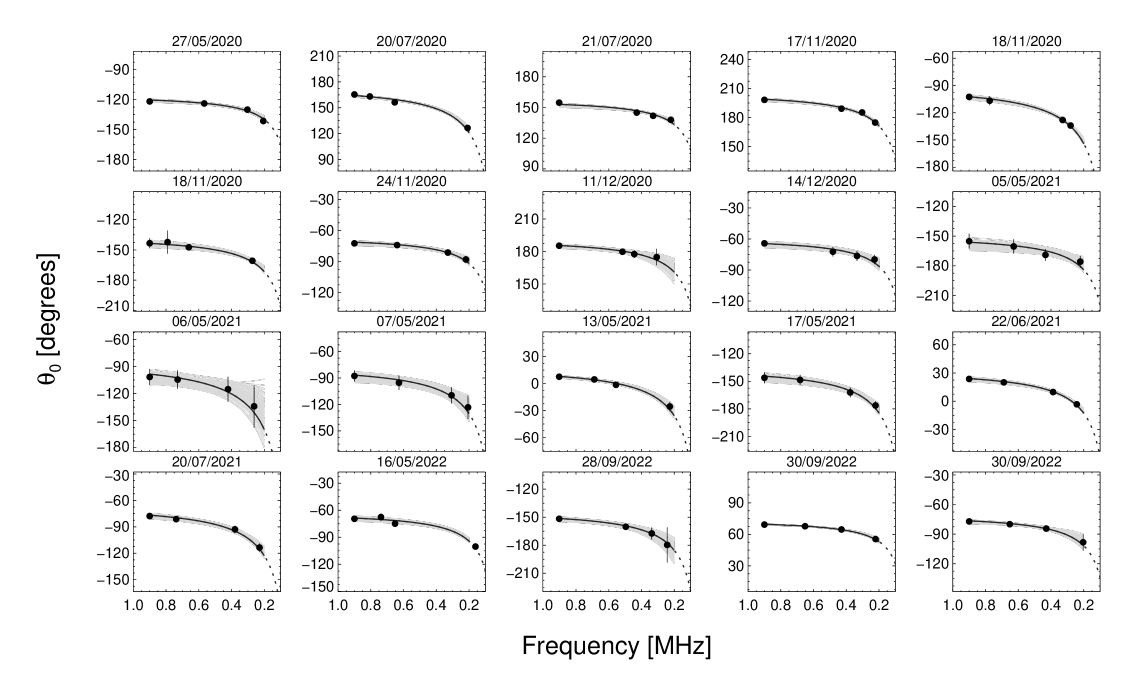}
   \caption{Heliocentric longitude $\theta_0 (f)$ of maximum intensity for twenty selected events. The fitted observed peak intensities (equation~\ref{eq:int_fit})) are given by black circles. The black curves show the best fits to $\theta_0$ using the model $\theta_0=\gamma_1/f - \zeta$. The grey regions show the ensemble of fits to the randomly perturbed data within the uncertainty of the individual values of $\theta_0$.}
\label{fig:th0_fits}
\end{figure}
        
The intensity fitting method is dependent on the calibrated fluxes. To test the dependency on the flux values, the peak fluxes were perturbed by a random factor within the range ($0.5-2$) and were re-fitted for $100$~realisations for all events, finding no significant change of the fitted peak longitudes $\theta_0$. For example, the average of the $100$~perturbed fitted peak longitudes for the event of 2020~July~21 vary by $<1$~degree at each frequency from that shown in Figure~\ref{fig:ds_lc_peakflux}. Figure~\ref{fig:th0_fits} shows $\theta_0$ as a function of frequency for the twenty events. We take the best fit to the data, as well as uniformly randomising $\theta_0$ at each frequency within the uncertainty bounds $\theta_0 \pm \delta\theta_0$ of each value and repeating the fit for the each set of perturbed data $100$~times (grey regions). The equivalent fits are also performed in terms of distance, where the observed frequencies are mapped to radial positions using the density model presented in ref. \cite{2019ApJ...884..122K} and through $f_p \propto n_p^{1/2}(r)$, together with an assumption of either fundamental ($f=f_p$) or harmonic ($f=2 f_p$) emission. In frequency space, we fit $\theta_0 = \gamma_1/f - \zeta$, and in terms of distance, we fit $\theta_0 = \gamma_2 \, r - \zeta$.
   
\begin{figure}[ht]
   \centering
    \includegraphics[width=0.9\textwidth]{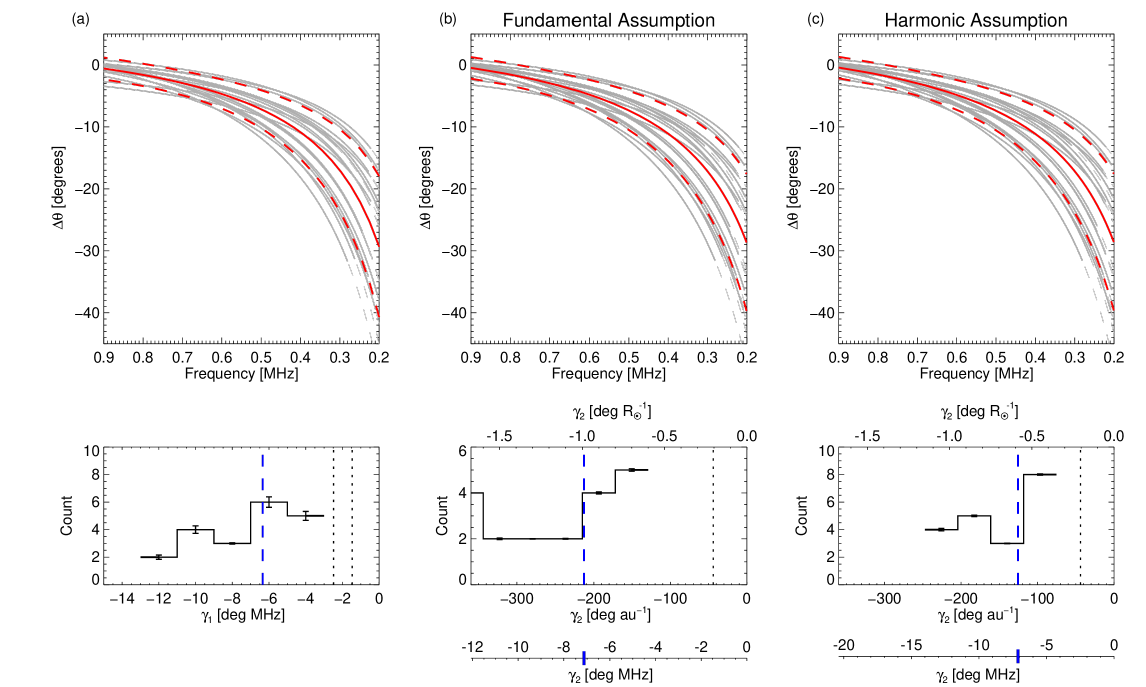}
    \caption{Individual lines of best fit to $\theta_0$ from
    Figure~\ref{fig:th0_fits}, normalised to $\theta_0$ at $0.9$~MHz using three different fitting approaches. Column~(a) shows the fits in frequency space from Figure~\ref{fig:th0_fits}. Columns~(b) and~(c) show the fits in terms of distance, for both fundamental and harmonic sources, converted back to frequency space for comparison. The solid red curves show the average of all the randomly perturbed fits down to $0.2$~MHz, with the dashed lines showing a standard deviation of the perturbed points. For events where the observed frequency at $0.2$~MHz was not available, the average was taken using the fit extrapolation (dashed lines). The lower panels show histograms of the fitted values of $\gamma_1$ and $\gamma_2$, and their weighted average (blue dashed line). The floating axes on the centre and right plots map the fit gradients to units of deg~MHz, for comparison with the left panel. The blue tick maps to the dashed line in the middle panels. The dotted black lines represent the Parker spiral for $v_\mathrm{sw}=380$~km~s$^{-1}$ (the frequency panel displays this value for both the fundamental and harmonic assumptions).}
\label{fig:angle_shift}
\end{figure}

Figure~\ref{fig:angle_shift} shows the best fits from Figure~\ref{fig:th0_fits} in terms of $\Delta\theta$ (upper panels), as well as the equivalent fits in terms of distance using both fundamental and harmonic assumptions. Since there is also some uncertainty in our choice of reference longitude $\theta_0$, we propagate this uncertainty in $\Delta\theta$. This approach results in a spread in $\Delta\theta$ at $0.9$~MHz. The $x$-axis of the panels in columns (b, c) have been converted back to frequency space for direct comparison to column (a). The dotted lines show the extrapolation of each fit beyond the minimum considered frequency. From the collated fits to the perturbed data shown in Figure~\ref{fig:th0_fits}, we take the average (solid red line) and standard deviation (dashed red lines) of the resulting points normalised to $0.9$~MHz. The lower panels show the fitted parameters $\gamma_1$ and $\gamma_2$ from the lines of best fit. For the fundamental and harmonic assumptions, we find an average longitudinal drift with distance of $-210^\circ$~au$^{-1}$ and $-130^\circ$~au$^{-1}$, respectively. For comparison, the additional axis for the centre and right panels converts the units of $\gamma_2$ to that of $\gamma_1$ as $\gamma_2 \; [\mathrm{deg\;MHz]} = \gamma_2 \;[\mathrm{deg\;} R_\odot^{-1} ] \cdot \left[ d(1/f)/dr \right]^{-1}$, according to $f(r) \propto C \, r^{-1.15}$ for $f \propto n_p^{1/2}(r)$, with $n(r) \simeq 1.4 \times 10^6 \, r^{-2.3}$ above $10 \, R_\odot$ (see equation~(43) in ref. \cite{2019ApJ...884..122K}). The constant $C$ depends on whether fundamental ($C=1$) or harmonic ($C=2$) emission is under consideration. The weighted average of the gradients $\gamma$ are given by $\overline{\gamma}=\sum{\gamma_i w_i}/ \sum{w_i}$ with weights $w_i=1/(\delta\gamma)^2$, where $\delta\gamma$ is the uncertainty on $\gamma$ from the fitting procedure.
        
\subsection*{Radio-wave Propagation Simulations}\label{sec:rw_simulations}

\begin{figure}[ht]
   \centering
   \includegraphics[width=.6\textwidth]{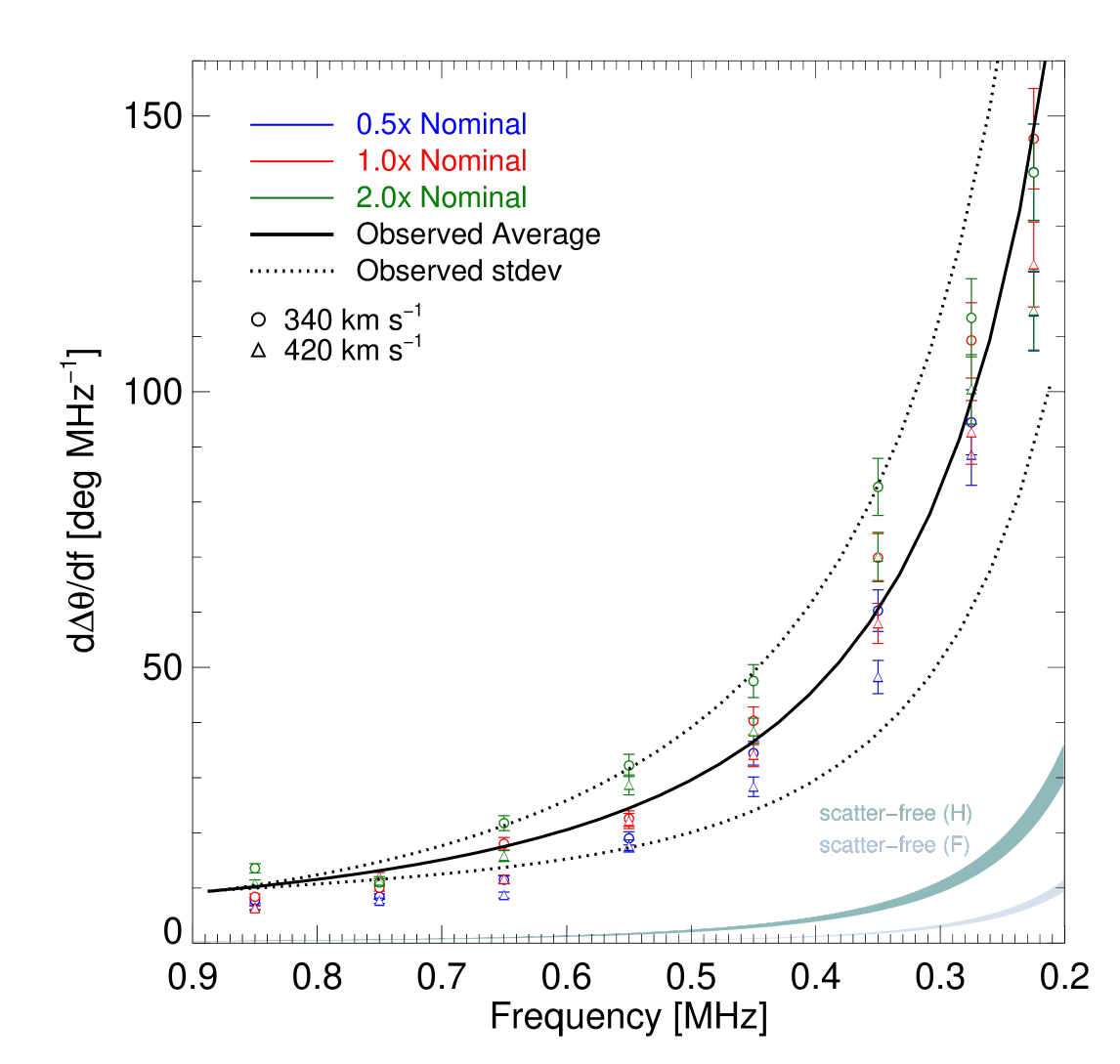}
   \caption{Simulated gradient $d(\Delta\theta)/df$ against
        frequency for the simulation parameters used in Figure~\ref{fig:dth}b. The open circles and triangles represent solar wind speeds of $340$ and~$420$~km~s$^{-1}$, respectively. The black lines give the gradient of the observed average and standard deviation, and the coloured bands give the gradient of the scatter-free cases. The uncertainty in the simulated
        gradient is a combination of the scatter (represented by the standard deviation $\delta\theta_0$) and $25$\% uncertainty in the frequency (i.e., $50$\% uncertainty in the density).}
\label{fig:sim_plots}
\end{figure}
        
The simulation code is described in ref. \cite{2023ApJ...956..112K} (see their equation~(14), describing a turbulence profile that is proportional to the scattering rate of radio waves and solar simulations in ref. \cite{2025ApJ...978...73C}). The polar components of the ambient magnetic field in the solar equatorial plane are described by the Parker spiral form that presents an average magnetic field observed in the inner heliosphere \cite{2023ApJ...946L..44S}
\begin{equation}\label{eq:parker_spiral}
    \begin{split}
        B_\mathrm{r} &= B_0 \left( \frac{r_0}{r} \right)^2 \\
        B_\theta &= - \, B_0 \left( \frac{r_0}{r} \right) \left( \frac{\Omega\left[r - 0.07 \, r_0 \right]}{v_{\mathrm{sw}}} \right) \, \sin{\phi} \,\,\, ,
    \end{split}
\end{equation}
where $r_0=1$~au, $\phi$ is the polar angle ($=\pi/2$ at the solar equator), $B_0 \simeq 5 \times 10^{-5}$~G is the magnetic field strength near $1$~au, $\Omega = 2.7 \times 10^{-6}$~rad s$^{-1}$ is the solar rotation rate, and we consider solar wind speeds between $340-420$~km~s$^{-1}$. The offset distance $0.07$~au ensures a radial magnetic field below an average Alfv\'en radius of $15 \, R_\odot$ \cite{2023SoPh..298..126C}. The anisotropy of the density fluctuations is set to $\alpha=0.25$ for the fundamental and $\alpha=0.4$ for the harmonic \cite{2023ApJ...956..112K}. The photons are injected as a delta function in both space and time at the radial distances defined by the density model used in ref. \cite{2019ApJ...884..122K}. To ensure that all sources lie on the same field line for a given $v_\mathrm{sw}$ despite different initial radii, the sources for a given frequency are offset in heliocentric longitude. To determine $\Delta\theta$ from the simulations (Figure~\ref{fig:dth}b) we use the weighted average directivity angle of the photon wavevector distribution (Figure~\ref{fig:summed_photons}). The simulated gradient $d(\Delta\theta)/df$ against frequency is shown in Figure~\ref{fig:sim_plots}.

\section*{Data Availability}

The observational data analysed in this study were sourced directly from NASA’s Space Physics Data Facility (SPDF) at \url{https://spdf.gsfc.nasa.gov/} for WIND, ST-A, and PSP. The data are publicly available through SPDF and can be accessed via their data services. Data for SolO were sourced from LESIA at \url{https://rpw.lesia.obspm.fr/rpw-data/}.

\bibliography{main}

\section*{Acknowledgements}

This work is supported by UKRI/STFC grants ST/T000422/1 and ST/Y001834/1. 
N.C. acknowledges funding support from the Initiative Physique des Infinis (IPI), a research training program of the Idex SUPER at Sorbonne Université. A.G.E. was supported by NASA's Heliophysics Supporting Research Program through grant 80NSSC24K0244, and by NASA award number 80NSSC23M0074, the NASA Kentucky EPSCoR Program, and the Kentucky Cabinet for Economic Development. V.K. was supported by the STEREO/WAVES and Wind/WAVES projects and by the NASA grant 19-HSR-19\_2-0143. The authors thank the PSP/RFS, SolO/RPW, STEREO/WAVES, and Wind/WAVES teams for making the data available. Solar Orbiter \cite{2020A&A...642A...1M} is a mission of international cooperation between ESA and NASA, operated by ESA. The FIELDS experiment on the Parker Solar Probe spacecraft \cite{2016SSRv..204....7F} was designed and developed under NASA contract NNN06AA01C.
This research has made use of the Astrophysics Data System, funded by NASA under Cooperative Agreement 80NSSC21M00561.
    
\section*{Author contributions statement}

D.L.C. performed the data preparation, calibration, implemented and analysed the data (both observational and modelling), produced the figures, and was responsible for the initial drafting of the manuscript and editing of the text. E.P.K. conceptualised the study, designed the numerical simulations, supervised the data analysis and calibration, wrote and edited the manuscript. N.C. supported with the interpretation of the results and editing of the manuscript. A.G.E. and N.L.S.J. contributed to the writing, commenting and editing of the manuscript. V.K.and A.V. contributed to the data calibration and checks. All authors contributed to the scientific discussion of the results and reviewed the manuscript. 

\section*{Additional information}

The authors declare no competing interests.

\end{document}